\documentclass[twocolumn,aps]{revtex4-1}
\usepackage{amsmath, amssymb}
\usepackage{graphics, graphicx}
\usepackage{enumerate}

\numberwithin{equation}{section}

\begin{document}

\title{Backreaction of Hawking Radiation on a Gravitationally Collapsing Star I:\\ Black Holes?}

\author{Laura Mersini-Houghton}

\affiliation{DAMTP, University of Cambridge, Wilberforce Rd., Cambridge, CB3 0WA, England \\ and Department of Physics and Astronomy, UNC Chapel Hill, NC 27599, USA.}

\date{\today}

\begin{abstract}
Particle creation leading to Hawking radiation is produced by the changing gravitational field of the collapsing star. The two main initial conditions in the far past placed on the quantum field from which particles arise, are the Hartle Hawking vacuum and the Unruh vacuum. The former leads to a time symmetric thermal bath of radiation, while the latter to a flux of radiation coming out of the collapsing star. The energy of Hawking radiation in the interior of the collapsing star is negative and equal in magnitude to its value at future infinity. This work investigates the backreaction of Hawking radiation on the interior of a gravitationally collapsing star, in a Hartle-Hawking initial vacuum. It shows that due to the negative energy Hawking radiation in the interior, the collapse of the star stops at a finite radius, before the singularity and the event horizon of a black hole have a chance to form. That is, the star bounces instead of collapsing to a black hole. A trapped surface near the last stage of the star's collapse to its minimum size may still exist temporarily. Its formation depends on the details of collapse. Results for the case of Hawking flux of radiation with the Unruh initial state, will be given in a companion paper $II$. 
\end{abstract}

\maketitle

\section{Introduction}

The backreaction of Hawking radiation onto a star collapsing into a black hole, is a long standing problem of major importance. It carries the tantalizing possibility that black holes may not form at all. Hawking radiation arises as particle creation from curved space-time quantum field theory.Therefore if its backreaction on the star's dynamics is significant, it would provide an example of the crucial role quantum effects play on strong gravitational fields such as those of massive imploding stars.

In 1939, Oppenheimer and Snyder \cite{os}, found that the ultimate fate of a spherically symmetric collapsing star is a collapse to a black hole. In 1974 and 1975, Hawking \cite{hawkingbh}, then Parker \cite{parkerbh}, found that quantum effects from the curved space time produced by the strong gravitational field of a star collapsing into a black hole, give rise to a thermal flux of particle production, known as Hawking radiation. 

Despite the efforts devoted to understanding the physics of black holes since Hawking's discovery of radiation, the main stumbling block in these efforts has been the issue of information loss paradox \cite{infoloss} stated in various forms. One of the most spectacular paradoxes stemming from the information loss mystery, is the recent 'firewall' problem raised in \cite{marolf}, then followed by the forecasting of \cite{hawkingweather}. 

Amidst all the puzzles and paradoxes, a trivial possibility is that black holes may not form. This possibility is the focus of the current study. Within a set of approximations, such as assumptions of spherical symmetry and homogeneity of a star collapsing into a black hole, this work shows that a black hole may not form when the backreaction of the quantum flux of particles created is taken into account in the collapse dynamics of the star.

In the next stage of this investigation, we drop some of the assumptions made here in solving the 4 dimensional Einstein equations for the interior of the collapsing star with a radiation flux, and use the Unruh vacuum as the initial state of the quantum field in the far past \cite{unruh}. Those results and their comparison to the method presented here will be given elsewhere \cite{lmhmp2}.

\section{The Star's interior metric}

Consider a spherically-symmetric, uniform density, perfect-fluid star, undergoing gravitational collapse. The stress energy tensor of the fluid is

\begin{equation} 
T^{\mu\nu}=\left(\varrho+p\right)u^{\mu}u^{\nu}+pg^{\mu\nu}
\end{equation} 

where $\varrho$ is the mass-energy density, $u^{\mu}$ the fluid 4-velocity, and $p$ is the isotropic pressure.
 Hydrodynamic equations for this fluid star follow from energy conservation : $T^{\alpha\beta}_{;\alpha} = 0$, combined with the Einstein equations.

The metric in the interior of a spherically symmetric star, can be written in the general form

\begin{equation} 
ds_{int}^{2}=- e^{2\phi} dt^{2}+ e^{\lambda} dr^{2} + R(r,t)^{2} d\Omega^{2}
\label{misnermetric}
\end{equation} 

where $d\Omega^{2} = d\theta^{2} + sin^{2} \theta  d\phi^{2}$, and $R(r,t)$ is the area radius \cite{hmisner}. As a result of Birkhoff's theorem, in vacuum this metric should approach the Schwarzchild metric in the exterior of the star 

\begin{equation} 
ds_{ext}^{2}=-\left(1-\frac{2m}{r}\right)dt^{2}+\frac{dr^{2}}{(1-\frac{2m}{r})}+r^{2}d\Omega^{2}
\end{equation} 

 The set of dynamic constraints and hydrodynamic relations, from Einstein equations, $G_{\mu\nu}= 8\pi T_{\mu\nu}$ along with energy conservation, are given in \cite{hmisner}, \cite{lecturebh}. We briefly review here the relation of the above metric to the closed FRW universe that we use for the remainder of the paper, as well as the hydrodynamic equation for fluid's velocity. A relativisitc 'Lorentz' factor $\Gamma$ and a 'proper' fluid velocity $U$ are defined by the relation $\Gamma = D_{r}R$ and $ U = D_{t}R $, where the proper derivatives with respect to metric ~\ref{misnermetric} are given by $D_{t} = e^{-\phi} ( \partial_{t} )_{r}$ and $D_{r} = e^{-\lambda/2} ( \partial_{r} )_{t}$. With these definitions, it can be shown that {\it the above general metric can be written as a closed FRW universe for the case of any spherically symmetric homogeneous stars}. Consider $R(r,t) = \tilde{R}(r) A(t)$. The normalization of the star's fluid $4-$velocity $u_{\mu}u^{\mu} = -1$ yields the relation

\begin{equation}
\Gamma^{2} = 1 + U^{2} - \frac{2m}{R} =
 1 + R^{2} [ (\frac{\dot{A}}{A})^{2} - \frac{8\pi\varrho}{3} ]  
\end{equation}

Since the star is homogeneous, the terms inside the brackets depend on time only, thus they can be collected into a function $1/a(t)^{2}$, which brings the Lorentz factor into the form \cite{lecturebh}
\begin{equation}
\Gamma^{2} = 1 - \kappa \frac{r^{2}}{a(t)^{2}}
\end{equation}

where $\kappa = 1, 0,-1$ correspond to the usual closed, flat and open FRW cases, $ dot$ and $prime$ to time and radius derivatives, and without loss of generality we can take $\tilde{R}(r) = r$. The relation of $e^{\lambda}$ to $R, \Gamma$ is given by $\Gamma = D_{r}R = e^{-\lambda /2} \partial_{r}R$, while one of Einstein equations \cite{hmisner}  gives $\phi' =0$ for the case of homogeneous dust. Therefore the metric of Eqn. \ref{misnermetric} for the interior of the collapsing star becomes

\begin{equation}
ds^{2} = - dt^{2} + a(t)^{2} \left(\frac{dr^{2}}{1 - \kappa r^{2}} + r^{2} d\Omega^{2} \right) 
\end{equation}

The equivalence of spherically symmetric homogeneous stars to closed FRW universes provides the basis for the Oppenheimer-Snyder (OS) model that we sketch below next.

We wish to study the interior of the gravitationally collapsing star with a stress-energy tensor $T_{\mu\nu}$, including the backreaction of the thermal Hawking radiation, estimated for the Hartle Hawking vacuum,  with a stress-energy tensor $\tau_{\mu\nu}$ derived in \cite{candelashh}. The new hydrodynamic relations will now follow from the total energy conservation $ \left(T^{\mu\nu}  + \tau^{\mu\nu} \right)_{;\nu} = 0$ .

The dust case $p=0$ we consider here is a simplified model. In general even for homogeneous stars $D_{r}\varrho = 0$, the pressure term is radius dependent and self gravitating, thus important. In fact, even for dust $p=0$, modifications to the star, for example perturbations of the star's fluid or new terms in the metric,  may introduce pressure corrections, $\triangle p(r) \neq
  0$, that may artificially lead to caustics, shell crossings, or even singularities, although perturbations to the energy density may remain bound, $\varrho_{0} \rightarrow \varrho_{0} + \triangle \varrho$ at some finite radius. Therefore a detailed study of perturbations around the solutions given here is important, and will need to be done numerically. With this caveat in mind, let's turn to our system.

\subsection{Collapsing Star}
From Birkhoff's theorem applied to spherically symmetric objects, the exterior metric of a time dependent spherically star with mass $M$, can be written as
\begin{equation}  
ds_{ext}^{2}=-f(r)dt^{2}+\frac{dr^{2}}{f(r)}+r^{2}d\Omega^{2}
\end{equation}     
with $dr\Omega$ the solid angle and $f(r)=\left(1-\frac{2m}{r}\right)$, where the mass function $m(r,t)$ denotes the mass enclosed within the radius $r$. The effect of Hawking radiation on the exterior metric is negligible, thus the Schwarzchild metric is a good approximation to the exterior of the star.
As shown in the section above, during the collapsing phase, the interior of any spherically symmetric homogeneous star can be given by the metric of a closed FRW 'universe' ~\cite{lecturebh}.  Proper time is denoted by $\tau$ and conformal time by $\eta$ in this FRW metric:
\begin{multline}
ds_{in}^{2}=-d\tau^{2}+a^{2}(\tau)\left[d\chi^{2}+\sin^{2}\chi dr^{2}\right] \\
=a^{2}(\eta)\left[-d\eta^{2}+d\chi^{2}+\sin^{2}\chi dr^{2}\right]
\end{multline}
with $\tau = \int a(\eta)d \eta $, $r(\tau)= a(\tau)\sin \chi$.

\subsection{The Oppenheimer Snyder model}

Although idealized, the OS model is useful in capturing the features of the interior dynamics of dust stars and in highlighting the key steps we need below, for the matching procedure of the interior and exterior metrics at the surface of the star. The motion of the star's surface is needed for the derivation of particle creation by the collapsing star.

In the Oppenheimer-Snyder (OS) model ~\cite{os}, the surface of a gravitationally collapsing spherically symmetric star made up of dust with radius $R_s$, moves along
\begin{equation}
 R_{s}=\frac{1}{2}R_{s}(0)\left[1+\cos\eta\right]
\end{equation}
with scale factor $a=\frac{a_{0}}{2}\left[1+\cos\eta\right]$ and proper internal time $\tau=\left(\eta+\sin\eta\right)$.
The external Schwarzchild time is denoted by $t$. Matching at $t=0$, $R_{s}(0)=R_{0}$, gives the well known relations: $a_{0}=\sqrt{\frac{R_{0}^{3}}{2M}}$,  $\sin\chi_{0}=\sqrt{\frac{2M}{R_{0}}}$ and,

\begin{multline}  
t=2M\ln\left[\frac{\sqrt{\frac{R_{0}}{2M}-1}+\tan(\frac{\eta}{2})}{\sqrt{\frac{R_{0}}{2M}-1}-\tan(\frac{\eta}{2})}\right]\\
+2M\sqrt{\frac{R_{0}}{2M}-1}\left[\eta+\frac{R_{0}}{4M}\left(\eta+ sin(\eta) \right)\right]
\label{timeconversion}
\end{multline} 

The interior and exterior metrics of the star, Eq.~\ref{timeconversion}, are matched at the surface of the star by solving the following equation

\begin{equation} 
\label{last2} 
a^{2}(\eta)=f(r)\left(\frac{dt}{d\eta}\right)^{2}-\frac{1}{f(r)}\left(\frac{dr}{d\eta}\right)^{2}
\end{equation} 
 $R_{s}(\eta)=r(\eta)\mid_{r=R_{s}}=a(\eta)\sin\chi_{0}$ where $\chi_{0}$ corresponds to the radius of the star at its surface.
The external null cordinates ($u$, $v$) at the surface of the star, given from the usual relation to time and radius can, as a result of the matching of the two metrics, be expressed as functions of proper time in the interior

\begin{equation}
\label{last3} 
\begin{pmatrix}u\\
v
\end{pmatrix}
=t \mp r \mp \ln\left(\frac{r}{2M}-1\right)\mid_{r=R_{s}}= \begin{pmatrix}  
U_{s}(\eta) \\
V_{s}(\eta)
\end{pmatrix}
\end{equation} 
in terms of internal coordinates $ \begin{pmatrix}  U \\ V \end{pmatrix}$ by replacing $( t(\eta)$, $r(\eta) )$ in Eqns.~\ref{last2}, and by demanding $v = 0$ at the beginning of collapse, (or $\eta = 0$). The center of the star $r=0$ corresponds to $U-2R_{s}(\eta = 0) $. Eqn \ref{last3} leads to the known result

\begin{align} 
v &=  V_{s}(\eta)=\ln\left[\sin^{2}\left(\frac{\eta}{2}\right)+\left(\frac{R_{0}}{2M}-1\right)\cos^{2}\left(\frac{\eta}{2}\right) \right.\nonumber\\
&\qquad \left. {} +2\sin\left(\frac{\eta}{2}\right)\cos\left(\frac{\eta}{2}\right)\left(\frac{R_{0}}{2M}-1\right)^{\frac{1}{2}}\right] \nonumber \\
&\qquad +R_{0}\left[\cos^{2}\left(\frac{\eta}{2}\right)+\left(\frac{R_{0}}{2M}-1\right)^{\frac{1}{2}}\sin\left(\frac{\eta}{2}\right)\cos\left(\frac{\eta}{2}\right)\right] \nonumber \\
&\qquad +\eta \left(\frac{R_{0}}{4M}+1\right)\left(\frac{R_{0}}{2M}-1\right)^{\frac{1}{2}}
 \end{align} 

Once the motion of the star's surface is known, then the energy of particle creation can be estimated (using Eqns \ref{twoone} - \ref{twotwo}).

\section{Hawking Radiation inside the star}  

Particle creation is a generic feature of curved space-time quantum field theory ~\cite{pc}.
The time-changing gravitational field of imploding stars gives rise to quantum gravitational particle creation. For the case of stars collapsing to Black Holes, this process is known as Hawking radiation. The whole flux of particles is created from the time the collapse starts, up to the point when the horizon forms, with the very last photon becoming the horizon. From the moment of horizon formation onward, the surface gravity $\kappa$ of the black hole is nearly a constant and no radiation can escape from the Black Hole to future infinity, since by the definition of the horizon, photons are trapped by the horizon. 

Particle creation can be thought of as arising from the 'tidal' forces of the changing gravitational field near the surface of the collapsing star that 'rip apart' vacuum pairs of particles antiparticles ~\cite{hawkingbh,hh,parkerbh,daviesunruh,bhliterature}. The positive energy particles escape and travel to future infinity becoming part of Hawking radiation, while the negative energy particles fall inward in the star ~\cite{hh}, ~\cite{bhliterature, daviesunruh}. If $\Sigma_{0}$ and $\Sigma_{f}$ denote the 3-surfaces at the onset of collapse time and the end of collapse respectively, then the time lapse between the two surfaces is the time interval during which all of the Hawking radiation is produced, with the final photon $\gamma_f$ being aligned to the horizon. Any other photon produced after $\gamma_f$ has to be trapped by the horizon and can not escape the black hole. 

The crucial point to be emphasized here is: Hawking radiation is produced by the changing gravitational field of the collapsing star, i.e. prior to the black hole formation,  ~\cite{daviesunruh}, ~\cite{pc}. Otherwise the surface gravity of the black hole $\kappa$, and the temperature of Hawking radiation would increase with time, leading to a nonthermal distribution of radiation. The event horizon of the black hole traps consequent photons of radiation produced near the event horizon. The photons produced after $\gamma_f$ can not travel outwards, rather their geodesics focus inside the black hole. More explicitly, the surface gravity of the Black Hole is defined in terms of the 4-acceleration of an external observer. If $\kappa$ were increasing with time, so would the acceleration of inertial relative to freely falling observers. For these reasons, quantum gravitational particle creation occurs during the collapse phase of the star,(see a seminal paper by Davies ~\cite{daviesunruh} for the details). 

Knowing that particle creation occurs during the collapse stage, means that we can include the backreaction of Hawking radiation onto the collapse dynamics of the star to find out  if a singularity forms at the end of the collapse. Thia is what we do next. We will take into account the backreaction of radiation on the star for the case when the initial state is in the Hartle-Hawking vacuum. The energy component of the star now receives a contribution from the negative energy flux of Hawking radiation that goes inside the star : $\varrho \rightarrow \varrho_{0} -|\varrho_{rad}|$, where $\varrho_{rad}$ is the energy density of Hawking radiation. Meanwhile, the pressure component in the stress energy tensor of the star will receive an additional contribution from this radiation which, based on the calculation of ~\cite{candelashh} for the Hartle-Hawking vacuum in 4 dimensions is given by : $\triangle p_{r} = \frac{\varrho}{3}$. The reason for the negative sign is that as explained above, thinking of Hawking radiation as a pair-creation of particles near the collapsing star, is such that the positive energy particles escape to infinity, but the negative energy particles fall inside the star ~\cite{hawkingbh,parkerbh,hh}. The negative energy Hawking radiation inside the star is a way of explaining the decrease in the star's mass with time. The radiation is time dependent during the collapse but, in the absence of backscattering, it is not radius dependent ~\cite{candelashh}. Considerable literature is devoted to this subject ~\cite{bhliterature}. 

In particular, one may worry that since radiation is actively being produced as the star collapses while at the same time backreacting on the star's metric, whether the time dependence of this flux of radiation is significant to the isotropy and symmetries of the star. We can address this concern by the following analogy in which the time dependence is known: the amount of Hawking radiation produced by a star collapsing into a black hole is equivalent to radiation produced by a moving mirror with a particular trajectory $\alpha(u)\sim\ln\cosh\left(\frac{\tau}{M}\right)$, shown in ~\cite{movingmirror}.
Therefore an elegant way of including the time dependence of Hawking radiation backreacting on the interior of the star in our calculation here, is to think of our system as a star undergoing the usual gravitational collapse but which has a mirror moving along with its surface.

At this stage we are not concerned about the fine details of the star backreacting on the mirror although once the modified collapse of the star is known then estimating the spectrum of particle creation is straightforward ~\cite{daviesunruh}, and we provide this result in the next section. It is the end state of the collapsing star we are interested in. Without loss of generality, we will keep the mirror along the trajectory which produces the right expression for the Hawking radiation, although the surface of the star will move along a modified path of collapse due to the backreaction of the Hawking radiation. Then, Hawking radiation is estimated from the modified motion of the star's surface, shown next. 

There are two ways to proceed from this point in solving the problem of an imploding star in a bath of negative energy radiation.
\begin{enumerate}[i]
            
\item We can exploit the symmetries of the problem and consider a Schwarzchild metric in the exterior and a "closed universe" in the interior. We then solve the Einstein equations in the interior of the star with the backreaction of Hawking radiation included, and match the two metrics at the moving surface of the star. The matching at the star's surface, allows us to express the evolution of the star in terms of external observers. The latter is what a normal external observer would see for an observer co-moving along the surface of the star. Once the modified motion of the star's surface is calculated by these steps, we can then proceed to estimate the modified Hawking radiation arising from it;

\item Or, we can solve the 4 dimensional Einstein equations, including the flux of radiation, $\tau_{\mu \nu}$, from :  $G_{\mu \nu}=8 \pi ( T_{\mu \nu} + \tau_{\mu \nu})$,  in the interior of the star, for the case when the initial state is the Unruh vacuum.
\end{enumerate}

The main difference between the two methods is in the choice of the initial vacuum state. The choice of the Hartle-Hawking vacuum \cite{hh} leads to an idealized situation of a thermal bath of radiation which is time symmetric and finite everywhere ($\tau_{tr} = 0$). The choice of the Unruh vacuum breaks the time symmetry and results in a flux of positive energy radiation escaping from the star to future infinity and an equivalent but negative energy flux of radiation ($\tau_{tr} \ne 0$) falling into the collapsing star \cite{candelasunruh, israel}. There is a vast ammount of literature devoted to a discussion of these two vacua states ~\cite{bhliterature,daviesunruh,unruh,hh,parkerbh}. Concerns with the first choice stem from the idealization of time symmetry, while concerns with the Unruh's vacuum arise from the question whether radiation is real. Even without a star, a uniformly accelerated observer would detect and Unruh flux of radiation. Despite the discussions on the choice of initial vacua, the existence of Hawking radiation is not disputed and it remains so independent of the choice of boundary conditions ~\cite{bhliterature}. The study of the backreaction of Hawking radiation on the star's interior for both vacua states would complete this part of investigation. In case $(ii)$ the choice of Unruh's vacuum produces  a flux of radiation $\tau_{tr}$. Solutions to Einstein equations and hydrodynamic equations for this initial state, including the case when the homogeneity assumption for the star is relaxed, will be presented in paper II \cite{lmhmp2}. The results of case $(i)$, the backreaction on the interior of the star, of the thermal bath of particle comprising Hawking radiation when the initial state is the Hartle-Hawking (HH), are presented here. 
Due to the warped nature of the metric in the star's interior ~\cite{visser1}, and a relation of the total stress energy tensor to Kodama flux ~\cite{kodama}, it will be show in ~\cite{lmhmp2} that the Hawking radiation backreaction effect on the star's interior leads to the same conclusions.

\subsection{Hawking Radiation in 4-dimensions}
\label{sec:Hawk}  

Authors of ~\cite{daviesunruh} investigated particle creation on the Schwarzchild barrier. They showed in the 2-dimensional case that the amount of negative radiation going into the star is equal to its counterpart Hawking radiation at future infinity. Davies further showed in ~\cite{movingmirror} that the amount of Hawking radiation at $\infty$, is equal to the radiation produced by a moving mirror with a trajectory r(t), such that $U= \alpha (u) = 2t_{u}-u$, $t-r(t_{u})=u$ where lower case ($u$,$v$) denote null coordinates in the star's exterior metric, and ($U$,$V$) in its interior metric. In general an interior metric in 2 dimensions is conformally flat and can be written as 
\begin{equation}
\label{twoone}
ds^{2}=C(U,V)dUdV= A(u,v)dudv
\end{equation}

Here $A(u,v) = C(U,V)\frac{DU}{du}\frac{DV}{dv} $.
Particle creation from this metric has a stress-energy tensor ~\cite{daviesunruh} 
\begin{equation}
\label{twotwo}
\begin{cases}
\tau^{finite}_{uu} & = -F_{u}(C) \\
\tau^{finite}_{vv} & = -F_{v}(C)
\end{cases}
\end{equation}
where : $F_{x}(y)=\frac{1}{12\pi}\sqrt{y}\frac{\partial^{2}}{\partial x^{2}}\left(\frac{1}{\sqrt{y}}\right)$
Similar results hold in 4 dimensions ~\cite{candelashh, candelasunruh}.
Once we transform the interior metric coordinates in Eq. \ref{twoone} in terms of the exterior metric ones, then Hawking radiation is derived from \ref{twotwo}. The different values of \ref{twotwo} correspond to different Boundary Conditions. For example Eq.~ \ref{twotwo}, for the choice of Boulware vacuum, corresponds to  the vacuum polarization term of a static star; the choice of the Unruh vacuum corresponds to a flux of particles ($\tau_{t r} \neq 0$ ) which is finite at $+\infty$, but divergent on the past horizon; the choice of Hartle-Hawking (HH) vacuum used in our investigation here, corresponds to a bath of particles with negative energy in the star's interior, and positive energy particles on the exterior of the star escaping to null infinity. In the HH vacuum, $\tau_{\mu \nu} $ is finite everywhere including at the horizon, but $\tau_{tr}=0$ ~\cite{hh, israel, daviesunruh, bhliterature}.

The results of this process in 4-dimensions for the energy density $\varrho_{HH}$ and pressure $p_{HH}$ were derived by Howard and Candelas ~\cite{candelashh}, (who also demonstrated that the isotropy of radiation is not broken).
They showed that in the HH-vacuum, the stress energy tensor of Hawking radiation $\tau_{\mu \nu}$ is such that : $\varrho_{HH} = 3 p_{HH} $, at $+\infty$ at future infinity, and equal and opposite to this value in the interior of the horizon.
This means that infalling particles at the horizon as well as those at $+\infty$, will not break the isotropy of the star. Let us now imagine a collapsing star filled in its interior with negative energy but isotropic thermal Hawking radiation. Due to the equivalence of the expression of Hawking radiation with the energy of particles produced by a moving mirror in a specific trajectory ~\cite{movingmirror}, let us take the time dependent energy of radiation being produced $\rho _{HH}$ given by the moving mirror with trajectory asymptotically approaching $u$, $\alpha(u)= t_{u}\ln\cosh\left(\frac{t}{M}\right)$ with $t$ proper external time. An estimation of the time-dependent renormalized stress energy tensor of particles radiated by the mirror in the interior of the star via: $\tau_{vv}^{Renorm}=F_{v}(\alpha)=\frac{\left(1+e^{-2\frac{t}{M}}\right)}{48\pi M^{2}}\left(-1+\frac{2}{1+e^{2\frac{t}{M}}}\right)\underset{{\scriptscriptstyle {\scriptstyle {\scriptscriptstyle \tau\rightarrow\infty}}}}{{\displaystyle \rightarrow}}-\frac{1}{48\pi M^{2}}$, which agrees perfectly with the energy of Hawking radiation in the HH-vacuum found in ~\cite{candelashh}, for the 4-dimensional renormalized stress-energy tensor.

The calculation is organized as follows: our collapsing star is filled with a Hawking negative energy thermal bath of radiation with 
the expressions for $\varrho_{HH}= 3 p_{HH}$ 'mimicked' by the moving mirror expression above.
We solve for the interior metric of the star, including the back reaction of the 'mirror' ($\varrho_{HH}$,$p_{HH}$). It is useful to recall that Hawking radiation doesn't depend on radius and on the details of the collapse. Then the internal metric is expressed in terms of the Schwarzchild exterior metric by matching at the moving surface of the star. The motion of the star's surface, determined by solving Einstein equations with the backreaction term included, allows us to estimate the modified Hawking radiation through Eqn. \ref{twotwo} ~\cite{movingmirror}, namely: for the new metric, we estimate $C^{New}(U,V)$ at the star's surface. We then have: $\tau^{New}_{vv}=-F_{v}(C^{New})$.
This method is equivalent to solving Einstein equations, since the pressure equation above related to the curvature of the star's metric, ($\frac{\ddot{a}}{a}=-\left(\varrho+\rho\right)$), is nothing more than the covariant energy conserved, which is taken into account. We then find that the star bounces at a finite radius instead of reaching the singularity.

\section{Backreaction of Hawking Radiation on the collapsing star}
Let us start with the choice of the HH-vacuum here, as the boundary condition at past infinity for studying the gravitational collapse dynamics in the interior of the star ~\cite{daviesunruh, candelashh,israel}. We use the matching of metrics at the surface of the star, illustrated in $II.B$ for the OS model, within the set of approximations made here, namely: assume a collapsing spherically symmetric dust star which is homogeneous and isotropic: $\varrho = \varrho_{0}=\frac{M}{\frac{4}{3}\pi R_{0}^{3}} $ , $p=0$. With the choice of the HH vacuum, this star is immersed in the thermal bath of Hawking radiation, which in the interior is given by the stress energy tensor $\tau_{\mu\nu}$ with components  $- 3\varrho_{HH} = - p_{HH}$, as derived in ~\cite{hh,candelasunruh,daviesunruh}.

The interior of the spherically symmetric star can be written as a closed FRW universe, as shown in Section II, for as long as we match the FRW metric at the surface of the star to the Schwarzchild metric ~\cite{lecturebh}. With the given metric in the interior, we can now solve Einstein equations. The $G_{00}$ component of Einstein equations is the "Friedmann" equation of the 'closed FRW universe' star.
\begin{equation} 
\label{closedu}
\left(\frac{\dot{a}}{a}\right)^{2}=\left(\frac{a'}{a^{2}}\right)^{2}=-\frac{1}{a^{2}}+\frac{\varrho_{0}}{a^{3}}-\frac{|\varrho_{rad}|}{a^{4}}
\end{equation} 
with $0 \leq \eta \leq \pi$, where:  $\dot{} = \frac{\partial}{\partial\tau}$, and $'=\frac{\partial}{\partial\eta}$. Here $\varrho_{0}$ is the homogeneous star's density, and $\varrho_{rad}$ is the Hawking radiation energy density in the interior of the star, which is negative.
The solution to Eqn \ref{closedu} is:
\begin{equation} 
\label{solu}
a(\eta)=\frac{\varrho_{0}}{2}-\frac{1}{2}\sqrt{\left(\varrho_{0}^{2}-4|\varrho_{rad}(\eta)|\right)\sin^{2}\left(\eta\right)}
\end{equation} 
Note that this solution \ref{solu} takes into account the time-dependence of $\varrho_{rad}(\eta)$ estimated in Section \ref{sec:Hawk} through the analogy with the moving mirror on the surface of the star. We can include that dynamics by now replacing $\varrho_{rad}(t)\rightarrow\varrho_{rad}(\eta)$ once we relate the exterior time $t$ to conformal interior time $\eta$ through the matching of the two metrics at the surface of the star .

The flow lines of the dust in the star's interior are determined by the implosion rate of the star's metric in the interior, $a(\eta)$, since $r = sin(\chi) a(\eta)$. So, the radial component of the fluid 3-velocity goes as $ v\simeq \dot{r} = sin(\chi) \dot{a(\eta)}$ implying that should the star bounce, the bounce conditions is given by $\dot{a}(t) = 0 $ 

{\it From Eqn. ~\ref{closedu} it is now clear that there always exists a solution for which there is a minimum radius the star reaches before it bounces}, $\dot{a}(t)=a'(\eta)=0$. This solution is roughly given by $a_{min}(\eta)\simeq\frac{\varrho_{rad}}{\varrho_{0}}$. Independently of the amount of Hawking radiation in $\varrho_{rad}$ or the fine details of the time dependence of $\varrho_{rad}$, the bounce condition is always satisfied for a finite radius.

The modified motion of the surface of the star given by $R_{s}(\eta)= a(\eta) sin(\chi_0)$, with $a(\eta)$ estimated in ~\ref{solu}, introduces corrections to the particle creation spectrum \ref{twotwo}.
To complete the calculation, if we are interested to find how Hawking radiation itself is modifed from the backreaction of the star's modified collapse, we first need to relate the interior conformal time $\eta =\eta(t)$ to the external time $t$ in order to express $\varrho_{HH}(t)= \varrho_{rad}$ of Section III.A as a function of internal conformal time $\eta$

\begin{equation}
\label{rad1}
\varrho_{rad}(t)=-\frac{1}{48\pi M^{2}}\left(1+e^{-\frac{t}{M}}\right)\left(-1+\frac{2}{1+e^{\frac{t}{M}}}\right)
\end{equation}
The conversion to interior conformal time is now achieved by matching the exterior and interior metrics at the moving surface of the star using Eqn.~\ref{last2}. The expression $\eta =\eta(t)$ we find is algebraically tedious. We used Mathematica to estimate it from Eqn.~\ref{last2} and approximate it with a simpler form
\begin{equation}
\label{rad2}
\varrho_{rad}(\eta)\sim-\tanh^{2}(\eta)\frac{1}{48\pi M^{2}} = - \tanh^{2}(\eta)\varrho^{0}_{rad}
\end{equation}
Although the actual expression for $\varrho_{rad}(\eta)$ for Eqn \ref{rad1} and $\eta(t)$ is messy, Eqn. \ref{rad2} fits it well and this is what is used in estimating the scale factor $a(\eta)$.

The second Einstein equation is trivially satisified: $\left(\frac{\ddot{a}}{a}\right)\sim-\Sigma_i (\varrho_i+ p_i)$ as it is equivalent to the covariant energy conservation (Bianchi identity), for both energy components $\varrho$ and $\varrho_{rad}$, already taken into account when evaluating $p_{rad}=\frac{1}{3} \rho_{rad}$ in the limit $r -> \infty$ \cite{candelashh}, with $p=0$.

Let us next evaluate the interior proper time from the conformal time through : $\tau=\int a(\eta)d\eta$ with $a(\eta)$ in \ref{solu}. Again using Mathematica, this function can be plotted to show that proper time in the interior is well approximated by
\begin{equation}
\label{rad3}
\tau=\simeq\frac{A}{2}\left(-1+\eta+\cos^{2}\left(\frac{\eta}{2}\right)\right)\simeq\frac{A}{2}\left(\eta-\sin^{2}\left(\eta\right)\right)
\end{equation}

Finally we calculate the exterior Schwarzchild time, $t$, in terms of $\eta$ via equation \ref{last2}, numerically. It is given roughly by
\begin{equation}
\label{rad5}
t \sim t_{0}\left[\eta+\cos(\eta)e^{-2\left(\eta+\cos\left(\frac{\eta}{2}\right)\right)}\right]
\end{equation}
From the relation $ r = a(\eta)\sin\chi$, we can construct the null coordinate $V_{s}$ at star's surface as: $R_{s}(\eta) = R_{0} a(\eta) = a(\eta) \sin\chi_{0}$, via
 
\begin{equation}
\label{rad4}
V_{s}(\eta)=t(\eta)-R_{s}-\ln\left|\frac{R_{s}}{2m}-1\right|
\end{equation}

The modified Hawking radiation resulting from the backreaction corrected motion of the star's surface is given by Eqn. \ref{rad4}. The contributing factor is $\triangle \tau_{v v} = -F_{v}(V_{s}(\eta))$. Writing $a(\eta)$ as a function of $t$ by using Eqn.\ref{rad5} in Eqn.\ref{solu} to 'translate' what an external observer would see for a normal observer located on the surface of the star.
These results, which take into account the backreaction of Hawking radiation on the interior metric of the collapsing star, provide all the information we need about the bounce radius, the motion of star's surface and the modified spectrum of particles. Backscattering is ignored throughout.

For our purposes the key point is that the condition $\dot{a}(\eta)=0$ gives the radius of the bounce for the star. This radius given by

\begin{equation}
\label{rad6}
a(\eta)^2 =\frac{\varrho_{0}}{2}\left[1-\sin(\eta)\left(1-4\frac{\varrho_{rad}}{\varrho_{0}^{2}}\tanh^{2}\left(\eta)\right)\right)\right]
\end{equation}

reaches a minimum at $a_{min} \sim \frac{\varrho_{rad}}{\varrho_{0}}$. The bounce is also observed by external observers, as given by the relation $a(\eta(t))= a(t)$. Within our approach presented here, no singularity and no event horizon have formed yet, since the star bounces at finite $(\eta,\chi,t)$ before the horizon is reached. The size of the star at its bounce is
\begin{equation}
R_{s}^{bounce} = a_{min}(\eta) \sin(\chi_0)\simeq \frac{\varrho_{rad}}{\varrho_{0}} \sin(\chi_0)
\end{equation}

We can equivalently deduce the bounce of the star and show that it is reached before the horizon fomrs, from the Einstein equations and the total energy conservation ~\cite{hmisner} of the original interior metric Eqn.~\ref{misnermetric}. One of the Einstein equations for the interior metric of the homogeneous dust star, Eqn.~\ref{misnermetric}, with the Hartle Hawking radiation in it, relates the velocity of the star's dust to the gravitational potential and the luminosity $L =4\pi R^{2} \varrho_{rad}$ of Hawking radiation
\begin{equation}
D_{\tau}U = - \frac{m}{R^{2}} - \frac{L}{R}
\end{equation}
Since we are focused in the interior of the star, where the Hawking radiation energy density is negative, we can write $\varrho^{0 inside}_{rad} = -q$. Integrating the above equation and recalling $\phi ' = 0, U = D_{\tau}R =\partial_{\tau} R$, gives the relation
\begin{equation}
U = R \frac{\dot{A}(\tau)}{A(\tau)} = ( \frac{2m}{R} -\frac{2m}{R_{0}} + 4\pi q R^{3} )^{1/2}
\label{u}
\end{equation}

Thus, from $\dot{R} = U = D_{\tau}R$ we get $\tau = \int \frac{dR}{U}$ which demonstrates, after integration, that the point $R=0$ is not reached within a finite proper time. The bounce radius is here estimated equivalently from the condition $\dot{R} = U = 0 $, yielding the same result as before.

The result for the bounce derived from the metric of Eqn.\ref{misnermetric} is useful in confirming that the bounce is reached before the Schwarzchild surfaces form. The condition for the formation of Scharwzchild surfaces is given by $\Gamma^{2} \le 0$ where $\Gamma$ is given by

\begin{equation}
\Gamma^{2} = 1 + U^{2} - \frac{2m}{R} 
\label{g}
\end{equation}

for our system of the dust star with negative energy radiation in it. We found from Eqn.~\ref{u} that $U^{2} = ( \frac{2m}{R} - \frac{2m}{R_{0}} + 4\pi q R^{3} )$ which, when replaced in the expression for $\Gamma$ in Eqn.~\ref{g}, gives 
\begin{equation}
\Gamma^{2} = 1 - \frac{2m}{R_{0}} + \frac{4\pi q R^{3}}{3} > 0
\end{equation}
 
Clearly this quantity is always positive.
We conclude that the star never enters the Schwarzchild surfaces, meaning the bounce occurs before the formation of an event horizon. The reason behind this result lies on the fact that the inclusion of negative energy radiation in the interior of the star, violates the energy condition of the Penrose-Hawking singularity theorem ~\cite{singularity}.

\subsection{Temporary trapped surfaces?}
We have shown that a collapsing star does not reach the point of collapsing all the way to a singularity in its center. Instead it bounces before the event horizon has a chance to form. But is there a trapped surface, anywhere, even if temporary? We can address this issue by looking at photons emitted at location $(\chi_{e},\eta_{e})$ in the star and finding if the area $A$ swept by them is such that 

\begin{equation}
\label{rad7}
\frac{dA}{d\eta}\le0
\end{equation}

where $A=\int \sqrt{g_{\theta\theta} g_{\phi\phi}} d\theta d\phi= \pi r_{e}^{2}(\eta)$
From \ref{rad7} and $r_{e}(\eta)=\sin\left(\chi_{e}\right)a(\eta_{e})$, $r>a$, the condition \ref{rad7} becomes

\begin{equation}
\label{rad8}
\frac{d\left[\sin\left(\chi_{e}\right)a\left(\eta_{e}\right)\right]}{d\eta}\leq0
\end{equation}

Note that photons in our FRW closed universe obey $\frac{d\chi}{d\eta}=\pm1$, or $\left(\chi-\chi_{e}\right)=\pm\left(\eta-\eta_{e}\right)$.
These null geodesics and Eq. \ref{rad8} lead to 

\begin{equation}
\label{rad9}
\pm\cos\left(\chi_{e}\right)a\left(\eta_{e}\right)+\sin\left(\chi_{e}\right)a'(\eta_{e})\leq0
\end{equation}

This is the condition for the formation of trapped surfaces, which can also be written as

\begin{equation}
\label{rad10}
\cos\left(\chi_{e}\right)a\left(\eta_{e}\right)\left[1+\tan\left(\chi_{e}\right)\frac{a'}{a}\right]\leq0
\end{equation}

Recall $a(\eta)>0$, but $\left(\frac{\dot{a}}{a}\right)<0$ for the collapsing to $a = a_{min}$-stage. Thus \ref{rad10} implies that in the collapse stage, photons that originate from a location $(\chi_e, \eta_e)$ in the star, which satisifies the relation ~\ref{rad11} below, will be temporarily trapped, if

\begin{equation}
\label{rad11}
\tan\left(\chi_{e}\right)\left|\frac{a'}{a}\right|\geq1
\end{equation}

For the stage $a'=0$,$a=a_{min}$, the condition for trapped surfaces of Eqn.\ref{rad11} can not be satisfied. The same is true for the expanding phase $a>a_{min}$, the trapped surface condition of \ref{rad10} can not be satisfied.
But for the collapse stage of \ref{rad11} there could be temporary trapped surfaces which depends on the details of collapse, as long as the collapse is at a rate such that $H=\frac{a'}{a^{2}}\geq\frac{1}{a\tan\left(\chi_{e}\right)}$.

\section{Conclusions}

We considered a spherically symmetric time dependent star undergoing gravitational collapse into a black hole. Including the backreaction of negative energy Hawking radiation in the interior of the star, leads to a behaviour whereby the star collapses to a minimum radius, then bounces before a black hole event horizon or singularity have a chance to form. Despite that, temporary trapped surfaces may form for photons located in the interior of the star, during the interval of collapse near its minimum size stage, but not beyond this stage. We provide the condition for the formation of temporary trapped surfaces. It is interesting to note that although Hawking radiation seems universal, in the sense that it does not depend on the details of the collapse, only on the mass of the star, the possible formation of the trapped surfaces do depend on the rate at which the collapse is occuring. At least this is the case in our simple model of a dust star. During the stage that a temporary trapped surface exists around the star, the star would appear the same as a black hole would to an external observer since information about the star is temporarily hidden behind the trapped surface.

What might go wrong with this picture? It is possible that the assumptions of symmetries and dust for the star are not realistic. Thus a perturbation around the solutions found here may make the star unstable to collapse. 
Further work remains to be done for the stability of the solutions found here, and for testing if these findings depend on the choice of the initial state for the Hawking radiation. Assuming homogeneity and isotropy allowed us to think of the star's interior as a closed FRW universe and conclude that it bounces. If these assumptions were dropped, then numerical solutions are needed for the 4 dimensional Einstein equations of the system of a collapsing star with a flux $\tau_{tr}$ of negative energy being absorbed by the star. We will present the latter in ~\cite{lmhmp2}. However, considering that for both HH and Unruh initial states, the degree of symmetry of the problem is quite similar, and that Hawking radiation remains the same independent of the initial state, then it is likely that bounce solutions of the collapsing star found here for the case of the HH initial state, will persist.

{\it Acknowledgements : }
 LMH is grateful to $DAMTP$, Cambridge University, for their hospitality during the time this work was done. LMH would like to thank L.Parker for valuable discussions related to particle creation and M.J.Perry for useful discussions and comments on the manuscript.


\end{document}